\documentclass[conference]{IEEEtran}
\usepackage{graphicx,epsfig,amsmath,amssymb,bm}
\usepackage{amssymb,balance}
\usepackage{amsmath}
\usepackage{amsthm} 
\usepackage{tikz}
\usepackage{amssymb}
\usepackage{algorithmicx}
\usepackage{algpseudocode}
\usepackage{url}
\newtheorem{theorem}{Theorem}
\def\LB{\left(}         
\def\RB{\right)}        

\newfont{\bbb}{msbm10 scaled 500}

\newfont{\bb}{msbm10 scaled 1100}

\newcommand{\RR}{\mbox{\bb R}}

\newcommand{\av}{{\bf a}}

\newcommand{\cv}{{\bf c}}

\newcommand{\fv}{{\bf f}}
\newcommand{\gv}{{\bf g}}

\newcommand{\lv}{{\bf l}}

\newcommand{\nv}{{\bf n}}

\newcommand{\pv}{{\bf p}}

\newcommand{\xv}{{\bf x}}

\newcommand{\zv}{{\bf z}}

\newcommand{\Am}{{\bf A}}
\newcommand{\Bm}{{\bf B}}

\newcommand{\Dm}{{\bf D}}

\newcommand{\Hm}{{\bf H}}

\newcommand{\Wm}{{\bf W}}

\newcommand{\thetav}{\hbox{\boldmath$\theta$}}

\usepackage{bbm}
\usepackage{subcaption}
\usepackage{tabularx}
\usepackage{multirow}
\usepackage{verbatim}
\usepackage[ruled,linesnumbered]{algorithm2e}

\newtheorem{problem}{Problem}

\newtheorem{definition}[theorem]{Definition}
\newcommand{\beqa}{\begin{eqnarray}}
\newcommand{\eeqa}{\end{eqnarray}}
\newcommand{\dsp}{\displaystyle}

\ifCLASSINFOpdf

\else

\fi

\graphicspath{{/Users/subhash/Dropbox/ADSC/Since_2017/Latex/MTD/Figures/}}

\begin{document}

\title{Moving-Target Defense for Detecting Coordinated Cyber-Physical Attacks in Power Grids}
\IEEEoverridecommandlockouts 
\author{
\IEEEauthorblockN{Subhash Lakshminarayana\IEEEauthorrefmark{1}, E. Veronica Belmega\IEEEauthorrefmark{2} and H. Vincent Poor\IEEEauthorrefmark{3}} 
\IEEEauthorblockA{\IEEEauthorrefmark{1}
School of Engineering, University of Warwick, UK \\
\IEEEauthorrefmark{2}
ETIS, Universit\'e Paris Seine, Universit\'e Cergy-Pontoise, ENSEA, CNRS, Cergy-Pontoise, France\\ 
\IEEEauthorrefmark{3}  Department of Electrical Engineering, Princeton University, Princeton, NJ 08544, USA \\
Emails: \IEEEauthorrefmark{1}subhash.lakshminarayana@warwick.ac.uk, \IEEEauthorrefmark{2}belmega@ensea.fr \IEEEauthorrefmark{3} poor@princeton.edu}
\vspace*{-0.5cm}
\thanks{
This work was supported in part by a startup grant at the University of Warwick and in part by the U.S. National Science Foundation under Grants  DMS-1736417 and ECCS-1824710.}}

\maketitle

\begin{abstract}
This work proposes a moving target defense (MTD) strategy to detect coordinated cyber-physical attacks (CCPAs) against power grids. A CCPA consists of a physical attack, such as disconnecting a transmission line, followed by a coordinated cyber attack that injects false data into the sensor measurements to mask the effects of the physical attack. Such attacks can lead to undetectable line outages and cause significant damage to the grid. The main idea of the proposed approach is to invalidate the knowledge that the attackers use to mask the effects of the physical attack by actively perturbing the grid's transmission line reactances using distributed flexible AC transmission system (D-FACTS) devices. We identify the MTD design criteria in this context to thwart CCPAs. The proposed MTD design consists of two parts. First, we identify the subset of links for D-FACTS device deployment that  enables the defender to detect CCPAs against any link in the system.  Then, in order to minimize the defense cost during the system's operational time, we use a game-theoretic approach to identify the best subset of links (within the D-FACTS deployment set) to perturb which will provide adequate protection. Extensive simulations performed using the MATPOWER simulator on IEEE bus systems verify the effectiveness of our approach in detecting CCPAs and reducing the operator's defense cost. 
\end{abstract}

\IEEEpeerreviewmaketitle

\section{Introduction}
Cyber threats against power grids are of increasing concern due to the deep integration of information and communication technologies (ICT) into grid operation. A recent real-world example was the December 2015 cyber attack against the Ukraine's power grid which resulted in large-scale outages that lasted several hours \cite{Ukraine2016:Analysis}. The attack was carried out by opening several transmission line circuit breakers and simultaneously blocking the information lines (e.g., telephone lines) to cover up the attacks. Such attacks have alerted us to a general class of attacks called the coordinated cyber-physical attacks (CCPAs).

As the name suggests, a CCPA consists of two components, namely, a physical attack and a  cyber attack. The physical attack involves disconnecting a transmission line, generator or transformer. On the other hand, a cyber attack involves manipulating the sensor measurements that are conveyed from the field devices to the control center, and has an effect of masking the physical attack. The attacker may readily launch such a cyber attack by exploiting the power grid's communication vulnerabilities \cite{Dragonfly}. CCPAs can have severe effects on the grid, since undetected line/generator outages may trigger cascading failures, and have received significant recent attention \cite{SoltanCCPA2015, LiCCPA2016, LiCCPA2018, DengCCPA2017}.

To defend against CCPAs, recent studies \cite{LiCCPA2016} and \cite{DengCCPA2017} have proposed  strategies based on securing a set of measurements (e.g., by encryption) or relying on measurements from known-secure phasor measurement units (PMU) deployed in the grid. 
However, power grids consist of many legacy devices whose life cycles can last several decades, and incorporating major security upgrades in these devices can be  quite expensive. Moreover, extensive research has  shown that PMUs themselves are vulnerable to false data injection (FDI) attacks, which can be launched by spoofing their GPS receivers \cite{ShepardGPS2012}.

In this work, we propose a novel defense strategy to detect CCPAs based on the technique of moving target defense (MTD). As in prior works \cite{SoltanCCPA2015, LiCCPA2016, LiCCPA2018, DengCCPA2017}, we only consider physical attacks that disconnect the transmission lines. We note that to craft an undetectable CCPA, the attacker must obtain an accurate knowledge of certain line reactances \cite{LiCCPA2016, DengCCPA2017}. The main idea of the proposed MTD defense in this context is to invalidate the attacker's prior acquired knowledge by actively perturbing of the grid's line reactance settings. This can be accomplished using distributed flexible AC transmission system (D-FACTS) devices, which are capable of performing active impedance injection and are being increasingly deployed in power grids  \cite{DFACTS2007}. The proposed MTD defense strategy has the potential to make it extremely difficult for the attacker to track the system's dynamics and gather sufficient information to craft undetectable CCPA. The main contributions of this work are as follows:
\begin{itemize}
\item First, we formulate the MTD design problem to defend against CCPAs and identify the MTD design criteria in this context.

\item We then propose a solution to the D-FACTS deployment problem  using a graph-theoretic approach. Our proposed solution identifies the minimum-sized subset of links for D-FACTS deployment which enables the defender to detect CCPAs against any transmission line. 
\item However, an MTD solution that involves perturbing a large number the branch reactances can be expensive due to the MTD's operational cost \cite{LakshDSN2018}. To reduce the operator's cost of defense, during the system's operational time, we use a game-theoretic formulation to identify the best subset of links to perturb that will provide adequate protection.  
\end{itemize}
Extensive simulations conducted using the MATPOWER simulator shows the effectiveness of our solution. Moreover, the results show that the game-theoretic approach significantly reduces the operator's defense cost.

\section{Prior Work}
\label{sec:Related_Work}
Power grid security has received significant interest in the past few years. In particular, FDI attacks against power grid state estimation have been extensively studied \cite{Liu2009, Sinopoli_MarketOp2011, KimPoorProtection2011}. While FDI attacks affect only the sensor measurements that are conveyed to the control center (and hence consist only of a cyber attack), recent research \cite{SoltanCCPA2015, LiCCPA2016, LiCCPA2018, DengCCPA2017} has studied CCPAs attacks, which as noted above consist of both cyber and physical components. CCPAs were first proposed in \cite{SoltanCCPA2015} based on disconnecting a set of transmission lines and blocking sensor measurements from the attacked area. However, the proposed cyber attack cannot completely mask the effects of the physical attack. Moreover, under some conditions, it was shown that the operator can recover the phase angles and detect the physical attack using information from outside the attacked zone \cite{SoltanCCPA2015}. On the other hand, \cite{LiCCPA2016, LiCCPA2018} and \cite{DengCCPA2017} proposed the design of cyber attacks that can completely mask the effects of the physical attack under different assumptions about the attacker's knowledge. Further, \cite{LiCCPA2016, LiCCPA2018} and \cite{DengCCPA2017} have also investigated defense against CCPAs relying on a subset of protected measurements, which however is vulnerable (see Section~I).

Recently, the concept of MTD has been applied to defend against FDI attacks \cite{Morrow2012, RahmanMTD2014, LakshDSN2018, LiuMTD2018}. In comparison to these works, we are the first to apply MTD for defense against CCPAs. Our analysis shows that MTD for defending against CCPAs requires the formulation of novel design criteria both in terms of D-FACTS placement as well as D-FACTS perturbation selection in comparison to aforementioned works. Finally, we note that while game theory has been used in the context of defense against FDI attacks \cite{EsmaTSG2013,SanjabTSG2016}, this work is the first to apply it in the context of MTD design in power grids.

\section{System Model}
\label{sec:Prelim}

\subsubsection*{Power Grid Model}
\begin{figure}[!t]
\centering
\includegraphics[width=0.4\textwidth]{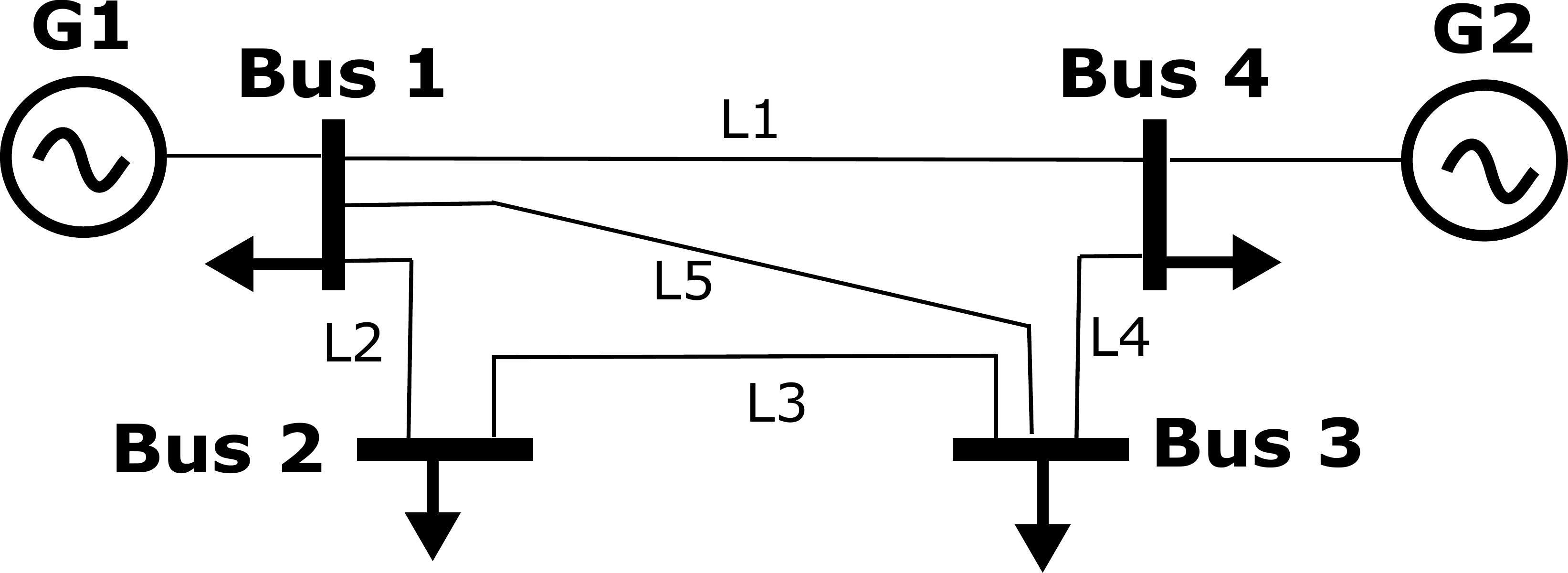}
\caption{An example of a $4$ bus power grid. }
\label{fig:4bus}
\vspace{-0.4 cm}
\end{figure}

We consider a power grid consisting of $N$ buses and $L$ transmission lines. The set of buses
and transmission lines are denoted by $\mathcal{N} = \{1,\dots,N\}$ and 
$\mathcal{L} = \{1,\dots,L\}$ respectively. An example of the IEEE-4 bus system with $5$ links is shown in Fig.~\ref{fig:4bus}.
At bus $i,$ we denote the amount of generation and load by
$G_{i}$ and $L_{i}$ respectively.  We let $l = \{ i,j\}$ denote a transmission line $l \in \mathcal{L}$ 
that connects bus $i$ and bus $j$ and its reactance by $x_{l}.$
The power flowing on the corresponding line $l$ is denoted by $F_{l},$ which under the DC power flow model \cite{wood1996power}
is given by $F_{l} = \frac{1}{x_{l}}(\theta_{i} - \theta_{j}),$ where $\theta_{i}$ and $\theta_{j}$ are the voltage phase angles at buses $i,j \in \mathcal{N}$ respectively. In vector form, the power flow vector  $\fv = [F_{1},\dots,F_{L}]^T$ is related to the voltage phase angle vector  $\thetav = [ \theta_1,\dots,\theta_N] $ as  $\fv = \Dm \Am^T \thetav,$ where the matrix $\Am \in \RR^{N \times L}$ is the branch-bus incidence matrix \cite{wood1996power} and $\Dm \in \RR^{L \times L}$ is a diagonal matrix of the reciprocals of link reactances.
We denote the set of links on which D-FACTS devices are deployed by $\mathcal{L}_D$ where $\mathcal{L}_D \subseteq \mathcal{L}.$ D-FACTS devices enable the reactances of these lines to be varied within a pre-defined range $[\xv^{\min} , \xv^{\max}],$ where $\xv^{\min},\xv^{\max}$ are the reactance limits achievable by the D-FACTS devices.

\subsubsection*{Optimal Power Flow}
For any given load condition $\lv = [L_1,\dots,L_N]$, the system operator sets the generation dispatch and line reactance settings by solving the optimal power flow (OPF) problem, 
stated as follows:
\begin{subequations}
\label{eqn:OPF_normal}
\beqa
 C_{\text{OPF}}  =  & \dsp \min_{ \gv,\xv} &  \sum_{i \in \mathcal{N}} C_i (G_{i}) \label{eqn:OPF_normala}
    \\ 
& s.t. &  \gv - \lv  = \Bm \thetav, \label{eqn:OPF_normalb}
 \\
& & \fv \in \mathcal{F}, \gv \in \mathcal{G}, \xv \in \mathcal{X},  \label{eqn:OPF_normald}
\eeqa 
\end{subequations}
where $C_i(\cdot)$ is the generation cost 
at bus $i \in \mathcal{N}.$ Equation \eqref{eqn:OPF_normala} is the nodal power balance constraint, where the matrix $\Bm =  \Am \Dm \Am^T.$ Constraints \eqref{eqn:OPF_normald} correspond to the branch power flows, generator limits, and D-FACTS limits, respectively, where $\mathcal{F} = [-\fv^{\max},\fv^{\max}], \mathcal{G} = [\gv^{\min},\gv^{\max}]$ and $\mathcal{X} = [\xv^{\min}, \xv^{\max}]$ and $\fv^{\max}$ is the maximum permissible line power flow (i.e., the thermal limit) and $\gv^{\min},\gv^{\max}$ are the generator limits. We note that in the absence of D-FACTS, OPF optimizes over the generator dispatch values only.

\subsubsection*{State Estimation \& Bad Data Detection}
The system state, i.e., the voltage phase angles $\thetav,$ are estimated from the noisy sensor measurements using the state estimation (SE) technique. The sensor measurements, which we denote by ${\zv} \in \RR^{M },$ correspond to the 
nodal power injections, and the forward and reverse branch power flows, i.e. ${\zv} = [\tilde{\pv},\tilde{\fv},-\tilde{\fv}]^T$ and $M$ is the total number of measurements, where $M = N+2L.$ We denote the sensor measurement noises by a vector $\nv \in \RR^{M },$ which is assumed to follow a Gaussian distribution.
Under the DC power flow model, the relationship between $\thetav$ and $\zv$ is given by 
${\zv} = \Hm \thetav + \nv,$ where $\Hm \in \RR^{M \times N}$ is the system's measurement matrix given by
$\Hm = [\Dm \Am^T;-\Dm \Am^T;\Am \Dm \Am^T]$ ($[\Am;\Bm]$ denotes the row concatenation of matrices $\Am$ and $\Bm$).
The maximum likelihood (ML) technique is used for system state estimation \cite{wood1996power}. 
Under ML estimation, the estimate $\widehat{\thetav}$ is related to the measurements $\zv$ as
$\widehat{\thetav} = ({\Hm^T} \Wm \Hm)^{-1} {\Hm^T} \Wm \zv,$
where $\Wm$ is a diagonal weighting matrix whose elements are reciprocals
of the variances of the sensor measurement noise components.

After state estimation, a bad data detector (BDD) computes a quantity referred to as the residual, which we denote by $r$ as $r = ||\zv - \Hm \widehat{\thetav}||.$ A bad data alarm is flagged if the residual exceeds a predefined threshold  $\tau.$ The threshold is adjusted to ensure that the false positive (FP) rate
does not exceed $\alpha,$ where $\alpha > 0$ (usually a small value close to zero).

\subsubsection*{Undetectable False Data Injection Attacks}
We denote the FDI attack vector by $\av \in \RR^{M},$ which the attacker injects into the sensor measurements and the measurement vector with the FDI attack by $\zv^a$, given by $\zv^a = \zv+\av.$ It has been shown 
\cite{Liu2009} that an FDI attack of the form $\av = \Hm \cv,$ where $\cv \in \RR^N,$ remains undetected by the BDD.  Specifically, the probability of detection for such attacks is equal to the FP rate  $\alpha.$ We call these attacks undetectable FDI attacks.

\subsubsection*{Coordinated Cyber and Physical Attack}
While an FDI attack only modifies the sensor measurements, a CCPA attacks the grid physically followed by a coordinated FDI attack on the sensor measurements, as noted above. In particular, we consider physical attacks that disconnect a set of transmission lines, e.g., by opening the line circuit breakers. The physical attack will alter the power grid's topology and power flow, and the mismatch between the pre-attack (i.e., line disconnections) and post-attack measurements can generally be detected by the BDD. However, it has been shown that if the attacker injects a carefully-constructed coordinated FDI  attack on the sensor 
measurements, then the effect of the physical attack on the BDD residual can be completely masked \cite{DengCCPA2017}. Hence, the attack remains undetected by the BDD. 

Denote the set of links disconnected by the attacker under a physical attack by $\mathcal{L}_A.$
We use the subscript $``p"$ to denote the power grid parameters following the physical attack. It can be shown that the grid measurements post the physical attack are related to the pre-attack measurements by
 $\zv_p = \zv+\av_p,$ where $\av_p = \Hm \Delta \theta + \Delta \Hm \thetav_p,$ where $ \Delta \Hm$ is the change in the measurement matrix before and after the physical attack, given by, $\Delta \Hm = \Hm-\Hm_p.$ Reference \cite{DengCCPA2017} showed that in order to mask the effect of the physical attack and remain undetected by the BDD, the attacker must inject a coordinated FDI attack of the form $\av = \Delta \Hm \thetav_p.$

\subsubsection*{Knowledge Required to Launch a CCPA}
Next, we enlist the knowledge required by the attacker to construct an FDI attack of the form $\av = \Delta \Hm \thetav_p.$ Assume that the attacker disconnects a single branch $\mathcal{L}_A = \{l \}$ that connects buses $i$ and $j.$ 
It can be easily verified that $\Delta \Hm$ depends on the tripped branch 
 reactance $x_l$ only. Therefore, to construct the attack $\av = \Delta \Hm \thetav_p$, the attacker must obtain knowledge of the branch reactance $x_l$ and the difference in phase angles of the buses $i$ and $j$ following the physical attack, i.e., $\theta_{i,p} - \theta_{j,p}$ \cite{DengCCPA2017}. The knowledge of $\theta_{i,p} - \theta_{j,p}$ can be obtained by monitoring the line power flows following the physical attack as follows:
\begin{align}
 \theta_{i,p} - \theta_{j,p} = - \sum_{ m \in p^{k}_l }  x_{l_m} F_{l_m,p}, \label{eqn:ph_diff}
\end{align}
where $p^k_{l} $ is any alternative path between nodes $i$ and $j$ in the residual power network following the physical disconnections, i.e., $\mathcal{L} \setminus \mathcal{L}_A. $ Each path $p^k_{l}$ in turn is a collection of links $p^k_{l} = \{ l_{k_1}, l_{k_2}, \dots, l_{k_M}\}$ such that
$src(l_{k_1}) = i$ and $dst(l_{k_M}) = j,$ and $k_M$ is the number of links in the path  $p^k_{l}.$
We denote by $\mathcal{P}_{l} = \{ p^1_{l}, p^2_{l}, \dots, p^{K_l}_{l} \}$ a collection of all alternative paths between buses $i$ and $j,$ where $K_l$ is the number of such alternative paths. Note that the subscript $l$ denotes the disconnected link.

In the IEEE-4 bus example, assume that the attcker disconnects link~1. 
After the disconnection, there are two alternative paths between buses $1$ and $4,$ and hence, $K_1 = 2.$ These paths are given by
$p^{1}_l = \{ 2,3,4 \}$ with $k_1 = 3$ and  $p^{2}_l = \{ 5,4 \}$ with $k_2 = 2.$ The attacker can compute the phase angle difference between nodes 1 and 2 using \eqref{eqn:ph_diff} as
$\theta_{1,p} - \theta_{j2p}  = - \LB x_2 F_{2,p} + x_3 F_{3,p} + x_4 F_{4,p} \RB$ or,  $ \theta_{1,p} - \theta_{j2p}  = - \LB x_5 F_{5,p} + x_4 F_{4,p} \RB.$

In \eqref{eqn:ph_diff}, the attacker can obtain the knowledge of power flows $F_{l_m,p}$ by monitoring the line flow sensor measurements. On the other hand, the line reactances $x_{l_m}$ can be learned by monitoring the grid power flows over a period of time using existing techniques \cite{PoorBlind2013, LakshICASSP2018}. The attacker can also learn the reactance of the disconnected branch $x_l$ similarly.

\section{Moving-Target Defense for CCPAs}
\label{sec:MTD}
In this work, we propose a solution to defend the system against CCPAs based on the MTD technique. The main idea behind this approach is to periodically perturb the branch reactances of certain transmission lines to invalidate the attacker's acquired knowledge. Hence, an attack constructed using outdated knowledge of the system can be detected by the BDD. (The reader can refer to \cite{LakshICASSP2018} for practical guidance on how frequently the branch reactances must be perturbed.) In this section, we first formalize the MTD design problem to defend against CCPAs. The solution to the MTD design problem is presented in Section~\ref{sec:MTD_Soln}.  The details are presented next.

Recall  from \eqref{eqn:ph_diff} that to construct an undetectable CCPA, the attacker must acquire the following: (i) knowledge of the reactance of the tripped branch, $x_l,$ and (ii) knowledge of branch reactances in at-least  one alternate paths $p^k_{l}$
between the nodes $i$ and $j.$ 
Therefore, under MTD, the defender can thwart the CCPA by invalidating one of the two:
\begin{itemize}
\item[C1.]  Invalidate the attacker's knowledge of the tripped branch's reactance $x_l$. 
\item[C2.]  Invalidate the attacker's knowledge of at-least one of the branches in the path $p^k_{l}$ between nodes $i$ and $j.$
\end{itemize}
Note that the defender however cannot have prior knowledge of which link the attacker chooses to disconnect. Moreover, for a disconnected link 
$l,$ the defender has no way of knowing which path $p^k_{l} \in \mathcal{P}_{l}$ the attacker may have used to compute the phase angle difference $\theta_{i,p} - \theta_{j,p}$ as in \eqref{eqn:ph_diff}. Thus, the defender must invalidate the attacker's knowledge of the reactance of at-least one branch in every path $p^k_{l} \in \mathcal{P}_{l}.$ The defender must do so for every link $l \in \mathcal{L}$ (such that the attacker cannot launch a CCPA by disconnecting any link in the grid). 
Based on the arguments above, the MTD perturbation selection problem can be stated as follows:
\begin{problem}[MTD problem]
For each branch $l \in \mathcal{L},$ invalidate the knowledge of at-least one of the branches in $ \{ l \} \cup p^k_{l}, \ k = 1,\dots, K_l.$
\end{problem}
The MTD perturbation problem poses constraints on the D-FACTS deployment set  $\mathcal{L}_D,$ since a preliminary requirement to invalidate the attacker's knowledge of a branch reactance is the presence of a D-FACTS device on that link. Thus, $\mathcal{L}_D$ must be chosen in a way that it gives the defender the ability to protect every link $l \in \mathcal{L}$. A trivial solution is to deploy a D-FACTS device on every link of the power grid. However, a system operator may wish to minimize the number of D-FACTS devices installed in order to minimize the device deployment cost.

On the other hand, MTD perturbations incur an opeartional cost for the defender. Reference \cite{LakshDSN2018} characterized this cost in terms of the increase in OPF cost of the grid due to the MTD perturbations\footnote{Note that in the absence of MTD, the D-FACTS settings are adjusted to minimize the OPF cost as in \eqref{eqn:OPF_normala}. Thus, the MTD perturbations will increase the OPF cost, and the MTD operational cost in non-negative.}.
Perturbing the reactances of a large number of links may be expensive. Thus, at the system's operational time, the defender may wish to perturb the reactances of only a subset of links, which we denote by  $\mathcal{L}_{D_w},$ where  $\mathcal{L}_{D_w} \subseteq \mathcal{L}_D,$ such that the attacker cannot launch CCPAs against some specific links that are perceived to be important and vulnerable to attack.

In what follows, we provide solutions to both the aforementioned aspects of MTD design problem. 
Specifically, we first present an algorithm to find the D-FACTS deployment set $\mathcal{L}_D$ that satisfies the MTD design problem with a minimum number of devices based on a graph-theoretic approach. Subsequently, we present a solution to the problem of selecting a subset of links $\mathcal{L}_{D_w}$ for 
reactance perturbation at the operational time based on a game-theoretic approach.

\section{Solution to the MTD Design Problem}
In this section, we solve the MTD design problem formalized in Section~\ref{sec:MTD}. We first address the problem of finding the D-FACTS deployment set.
\label{sec:MTD_Soln}
\subsection{D-FACTS Deployment}
Our key observation to solve the D-FACTS deployment set problem is that each set of links $\{l \} \cup p^k_l, k = 1,\dots, k_M,$ forms a loop in the graph $\mathcal{G}.$ For example, in the 4-bus example in Figure~\ref{fig:4bus}, assuming that the attacker disconnects link 1, the links $\{1 \} \cup \{2,3,4 \} $ and $\{1 \} \cup \{4,5 \} $ form loops in the corresponding graph. If a DFACTS device is installed on a subset of links in the graph such that every loop in the network has at least one link with a D-FACTS device installed, then the attacker cannot launch an undetectable CCPA. 

In graph-theoretic terms, the problem is equivalent to removing a subset of links in the network such that the residual graph has no loops. For optimized deployment, $\mathcal{L}_D$ must be the minimum number of such links. If each link is assigned a weight of $1,$ then $\mathcal{L}_D$ must be a subset of links with minimum weight.

The set $\mathcal{L}_D$ can be found by solving the \emph{minimum weight feedback edge set problem in an undirected graph} \cite{BondyGraph1976}. The solution proceeds by finding the \emph{maximum weight spanning tree} (MWST). Specifically, let $\mathcal{L}_{\text{sptr}}$ be the 
MWST of the graph $\mathcal{G}.$ If D-FACTS devices are installed on the links $\mathcal{L} \setminus \mathcal{L}_{\text{sptr}},$ then the attacker cannot find a loop within the graph whose branches do not have a D-FACTS device installed. Equivalently, the attacker cannot launch an undetectable CCPA. 
Further, since the links in $\mathcal{L}_{\text{sptr}}$ form a maximum-weighted spanning tree, $\mathcal{L} \setminus \mathcal{L}_{\text{sptr}}$ are the links with minimum weight which can be disconnected. Equivalently, the links $\mathcal{L} \setminus \mathcal{L}_{\text{sptr}}$ are the minimum number of links that satisfy the D-FACTS desgin problem described in Section \ref{sec:MTD}. Thus, the D-FACTS deployment set $\mathcal{L}_D = \mathcal{L} \setminus \mathcal{L}_{\text{sptr}}.$

Consider the D-FACTS deployment set $\mathcal{L}_D$ chosen according to the above arguments. Assume that the defender perturbs the reactances of the set of links $\mathcal{L}_{D_w} \subseteq \mathcal{L}_D.$ Then, we have the following:
\begin{itemize}
\item A physical attack against a link $l $ can be detected by the BDD if the links in  $\mathcal{L}_{D_w}$ ensure that the conditions listed in Problem 1 are satisfied for that link. We will henceforth refer to such a link to being ``protected" under the MTD link perturbation set  $\mathcal{L}_{D_w}.$
\item Naturally, based on the arguments stated in this section, if $\mathcal{L}_{D_w} = \mathcal{L}_D,$ then all the links $l \in \mathcal{L}$ are protected from the physical attacks.  
\end{itemize}

\subsection{MTD Perturbation Selection Using Game Theory}
MTD perturbations incur an operational cost, and perturbing the reactances of a large set of links may not be cost effective. In this section, we answer the question of how to select the appropriate perturbation set $\mathcal{L}_{D_w}$. The main idea is to protect only a subset of  links from physical attacks depending on the operational state of the system, as well as the perceived threat to those links. This is approached using  a game-theoretic formulation. The details are presented next.

\subsubsection{Game Formulation}
We define the strategic interactions between the attacker and the defender as a two-player non-cooperative game. To formalize this, we define the game as a triplet $\mathcal{G} \triangleq \left( \{D,A\}, \{\mathcal{S}_D, \mathcal{S}_A\}, \{u_D, u_A\} \right)$
in which the components are: (i) the set of players $\{D,A\}$;  (ii) $\mathcal{S}_D$ and $\mathcal{S}_A,$ the sets of actions that  defender and attacker can take respectively; and  (iii) the payoffs of the players $u_k: \mathcal{S}_D \times \mathcal{S_A} \rightarrow \mathbb{R}$ for $k\in \{D,A\},$ where $u_k (s_D, s_A)$ measures the benefit obtained by player $k$ when the action profile that has been played is $s=(s_D,s_A)$.

We denote the attacker's and the defender's action sets by $\mathcal{S}_A  = \{ a_0, a_1, \dots,  a_{N_A-1} \}$ and $\mathcal{S}_D  = \{ d_0, d_1, \dots,  d_{N_D-1} \}$ respectively, where $N_A$ and $N_D$ are the cardinality of the sets $\mathcal{S}_A$ and $\mathcal{S}_D$ respectively. The attacker's action set is the subset of links it disconnects physically. We denote the set of links disconnected by the attacker under action $a_i$ by $\mathcal{L}_{a_i},$ where, $\mathcal{L}_{a_i} \subseteq \mathcal{L}, \ i = 0,1, \dots,N_{A}-1.$ 
The action $a_0$ corresponds to the case when the attacker does not attack any link. The defender's action is to select a subset of links within $\mathcal{L}_D$ whose reactances will be perturbed. We denote the set of links chosen by the defender under action $d_i$ by  $\mathcal{L}_{d_i},$ where, $\mathcal{L}_{d_i} \subseteq \mathcal{L}_D, \ i = 1, \dots,N_{D}-1.$
The action $d_0$ corresponds to the case when the defender does not perturb the reactance of any link.

Next, we characterize the attacker's and the defener's payoffs. The cost of damage due to the attack can be characterized as follows. If the attacker disconnects a link $l$ that is protected by the defender (due to the MTD perturbations), then the CCPA will be detected by the BDD, and the system operator can quickly restore the link to ensure that the attack does not result in any further damage. For instance, the defender can quickly restore the circuit breaker of the disconnected link to a closed position.  On the other hand, if the attacker disconnects a link that is not protected by the defender, then the CCPA will go undetected. The link disconnection will result in redistribution of power flows. Consequently, all the links on which the power flows exceeds the corresponding thermal limits will experience physical damage, and will get disconnected from the grid. In this case, the system operator will have to initiate load shedding in order to ensure that the attack does not result in further damage. (Herein, we assume that the BDD will detect the attack once additional links are disconnected, since the attacker's data injection will only mask the effect of disconnection of the first link.) We denote the cost of load shedding at bus $i$ by $C_{i,s} (L^s_{i}),$ where $L^s_{i} (\leq L_i)$ is the quantity of load that is shed.  We denote $\lv_s = [L^s_{1},\dots,L^s_{N}].$

Let $C_{\text{OPF}} (a_m,d_n)$ denote the OPF cost when the attacker takes an action $a_m$ and the defender takes an action $d_n.$ It can be computed as follows:
\beqa
 C_{\text{OPF}}({a_m,d_n})   =  & \dsp \min_{ \gv, \lv_s} &  \sum_{i \in \mathcal{N}} C_{i,g} (G_{i}) + C_{i,s} (L^s_{i}) \label{eqn:OPF_shed}
    \\ 
& s.t. &  \gv - \lv + \lv_s  = \Bm_{a_m,d_n} \thetav, \nonumber
 \\
& & \fv \in \mathcal{F}, \gv \in \mathcal{G}, \nonumber
\eeqa 
where $\Bm_{a_m,d_n}$ is given by $\Bm_{a_m,d_n} =  \Am_{a_m,d_n} \Dm_{a_m,d_n} \Am_{a_m,d_n}^T.$ Here, $\Am_{a_m,d_n}$ is the bus-branch connectivity matrix when the attacker and the defender choose actions $a_m$ and $d_n$ respectively. These quantities are computed as in Algorithm~1.
\begin{algorithm}
  \small
\SetAlgoLined
\KwData{$a_m,d_n$}
\KwResult{$C_{\text{OPF}}({a_m,d_n})$}
Set branch reactances to $\xv_{d_n}.$  \\
Set $\Am_{a_m,d_n} = \Am_{a_0,d_0}.$ \\
Solve \eqref{eqn:OPF_shed} to obtain $C_{\text{OPF}}({a_0,d_n})$.
\\
\eIf{attack is successful}{
Recompute power flows after removing the branches in $\mathcal{L}_{a_m}.$ \\
Monitor the branches for which the power flow exceeds the line capacity. Denote such links by  $\mathcal{L}^c_{a_m}.$ \\
Set $\Am_{a_m,d_n}$, $\Dm_{a_m,d_n}$ by removing the branches  $\mathcal{L}_{a_m}$ and $\mathcal{L}^c_{a_m}.$
Solve \eqref{eqn:OPF_shed} to compute $C_{\text{OPF}}({a_m,d_n}).$} {Set $C_{\text{OPF}}({a_m,d_n}) = C_{\text{OPF}}(a_0,d_n).$} 
\caption{\small Cost Computation}
\end{algorithm}

Based on the formulation above, the defender's payoff is given by
\begin{equation*}
u_D(s_D,s_A) = \left\{ 
\begin{array}{ll}
C_{\text{OPF}}(d_0, a_0) - C_{\text{OPF}}(s_D, a_0),   & \text{if} \ \mathcal{I}_S = 0  \\
C_{\text{OPF}}(d_0, a_0) - C_{\text{OPF}}(s_D, s_A),  & \text{if} \ \mathcal{I}_S = 1,
\end{array}\right.
\end{equation*}
and the attacker's payoff is 
\begin{equation*}
u_A(s_D,s_A) = \left\{ 
\begin{array}{ll}
0, &  \text{if} \ \mathcal{I}_S = 0 \\
C_{\text{OPF}}(s_D, s_A) - C_{\text{OPF}}(d_0, a_0),  & \text{if} \ \mathcal{I}_S = 1,
\end{array}\right.
\end{equation*}
where $\mathcal{I}_S$ is an indicator variable  to represent the success ($\mathcal{I}_S = 1$) or failure of an attack  ($\mathcal{I}_S = 0$).
Both players aim to choose their actions such that their own payoff is maximized and although the game is not a zero-sum game, we can see that the two players have contradictory objectives. The above payoffs can be explained as follows. First, $C_{\text{OPF}}(d_0,a_0)$ denotes the benchmark operating cost of the defender when none of the players takes an action to either disrupt or defend the system. The term $C_{\text{OPF}}(s_D, s_A) - C_{\text{OPF}}(d_0, a_0)$ denotes the the additional cost incurred by the defender and caused by a successful attack, when the attacker chooses $s_A$ and the defender chooses $s_D$; the defender's aim is to minimize this cost whereas the attacker wants to maximize it. The term $C_{\text{OPF}}(s_D, a_0) - C_{\text{OPF}}(d_0, a_0)$ represents the additional cost incurred by the defender for choosing an action $s_D$ against an unsuccessful attack $s_A$; the defender will seek to minimize this cost while neutralizing the attack. Of course, the benefit of the attacker if its attack fails is equal to zero.

\subsubsection{Solving the Game Formulation}
The game described above is discrete and finite. In such an interactive situation, a natural solution is the Nash equilibrium (NE), which is a stable state to unilateral deviation. Mathematically this is defined as:
\begin{definition}
A strategy profile $(s_D^*, s_A^*)$ is an NE for the game $\mathcal{G}$ if the following conditions are met:
$u_D(s_D^*, s_A^*) \geq  u_D(s_D, s_A^*), \ \forall s_D \in \mathcal{S}_D,
u_A(s_D^*, s_A^*) \geq  u_A(s_D^*, s_A), \ \forall s_A \in \mathcal{S}_A.$
\end{definition}
This means that neither player has any incentive to unilaterally deviate and will lose in terms of utility otherwise. This type of game may not have a pure NE solution but it always has at least one mixed-strategy NE  \cite{Tirole1991}, which is the NE of the extension of the game $\mathcal{G}$ to mixed strategies. It is defined as follows: $\widetilde{\mathcal{G}} \triangleq \left( \{D,A\}, \{ \Delta_D, \Delta_A\}, \{\tilde{u}_D, \tilde{u}_A\} \right)$.
The action sets of the extended game $\widetilde{\mathcal{G}}$ are the probability simplices of dimension $N_k$, $k\in\{D,A\}$:
$\Delta_k = \left\{ p_k \in \mathbb{R}^{N_k}_+ \left| \sum_{j=0}^{N_k} p_{k,j} =1 \right. \right\}$
where $p_k =(p_{k,0}, \hdots, p_{k,N_k-1})$ is the discrete probability vector of player $k$ such that $p_{D,j}$ and $p_{A,j}$ represent the probability of choosing the action $d_j$ by the defender and the probability of choosing the action $a_j$ by the attacker, respectively. The modified payoffs are simply the resulting expected payoffs following the randomization of play: 
\begin{equation}
\tilde{u}_k (p_D,p_A) = \sum_{j=0}^{N_D-1} \sum_{i=0}^{N_A-1} u_k(d_j, a_i) \ p_{D,j} \ p_{A,i}.
\end{equation}
The mixed NE can be defined similarly to the pure strategy NE. 
\begin{definition}
A mixed strategy profile $(p_D^*, p_A^*)$ is a mixed an NE for the game $\mathcal{G}$ if it is a NE for the extended game $\widetilde{\mathcal{G}}$ and the following conditions are met:
$\tilde{u}_D(p_D^*, p_A^*)  \geq  \tilde{u}_D(p_D, p_A^*), \ \forall p_D \in \Delta_D,$ and 
$\tilde{u}_A(p_D^*, p_A^*)  \geq  \tilde{u}_A(p_D^*,p_A), \ \forall p_A \in \Delta_A.$
\end{definition}

The mixed NE can be computed by using the Von-Neumann indifference principle \cite{Tirole1991}, which basically says that: i) player $k$ is rendered indifferent (in terms of its expected payoff) between its pure actions that are played at the NE with strictly positive probability, by the choice of the other's mixed action $p_{-k}$, for any $k\in \{D,A\}$; and ii) the actions that are not played at the NE (their probability equals 0 at the NE) give strictly lower payoffs than the ones that are played (see i)), for both players. Formally, this is stated in the following.
\begin{definition}
A mixed strategy profile $(p_D^*, p_A^*)$ is a mixed NE for the game $\mathcal{G}$ if it is an NE for the extended game $\widetilde{\mathcal{G}}$ and the following conditions are met:
\begin{enumerate}
\item both players are indifferent among their own pure actions that
are played with positive probability at the NE
\begin{eqnarray*}
\tilde{u}_D(d_j, p_A^*) & = & \tilde{u}_D(d_i, p_A^*), \ \forall d_j, d_i  \in \mathcal{I}^*_D,\\
\tilde{u}_A(p_D^*, a_j) & = & \tilde{u}_A(p_D^*,a_i), \ \forall a_j, a_i \in \mathcal{I}^*_A.
\end{eqnarray*}
\item the pure actions that result in strictly smaller payoffs are played with zero probability at the NE
\begin{eqnarray}
\tilde{u}_D(d_j, p_A^*) & < & \tilde{u}_D(d_i, p_A^*), \ \forall d_j \notin \mathcal{I}^*_D,  d_i  \in \mathcal{I}^*_D,\nonumber \\
\tilde{u}_A(p_D^*, a_j) & < & \tilde{u}_A(p_D^*,a_i), \ \forall a_j\notin \mathcal{I}^*_A, a_i \in \mathcal{I}^*_A \nonumber
\end{eqnarray}
\end{enumerate}
where the sets $\mathcal{I}^*_k \subseteq \mathcal{S}_k, \forall k$ denote the actions that are played with strictly positive probability at the NE: $\mathcal{I}^*_D =\{ d_j \in \mathcal{S}_D: p_{D,j}^* >0 \}$ and $\mathcal{I}^*_A =\{ a_j \in \mathcal{S}_A: p_{A,j}^* >0 \}$. 
\end{definition}

All defender's actions that are not in the set $d_j \notin \mathcal{I}^*_D$ have zero probability at the NE $p_{D,j}^*=0$ (they are not played at all at the NE) and the same goes for the attaker, all actions $a_j \notin \mathcal{I}^*_A$ have zero probability $p_{A,j}^*=0$ at the NE. Definition 3 provides a simple way to compute the mixed NEs by solving a system of linear
equations and checking some conditions, which we adopt in this work.

\section{Simulation Results}
In this section, we perform simulations to show the effectiveness of the proposed defense. All the simulations are carried out using the MATPOWER simulator.

\begin{figure}[!t]
\centering
\includegraphics[width=0.45\textwidth]{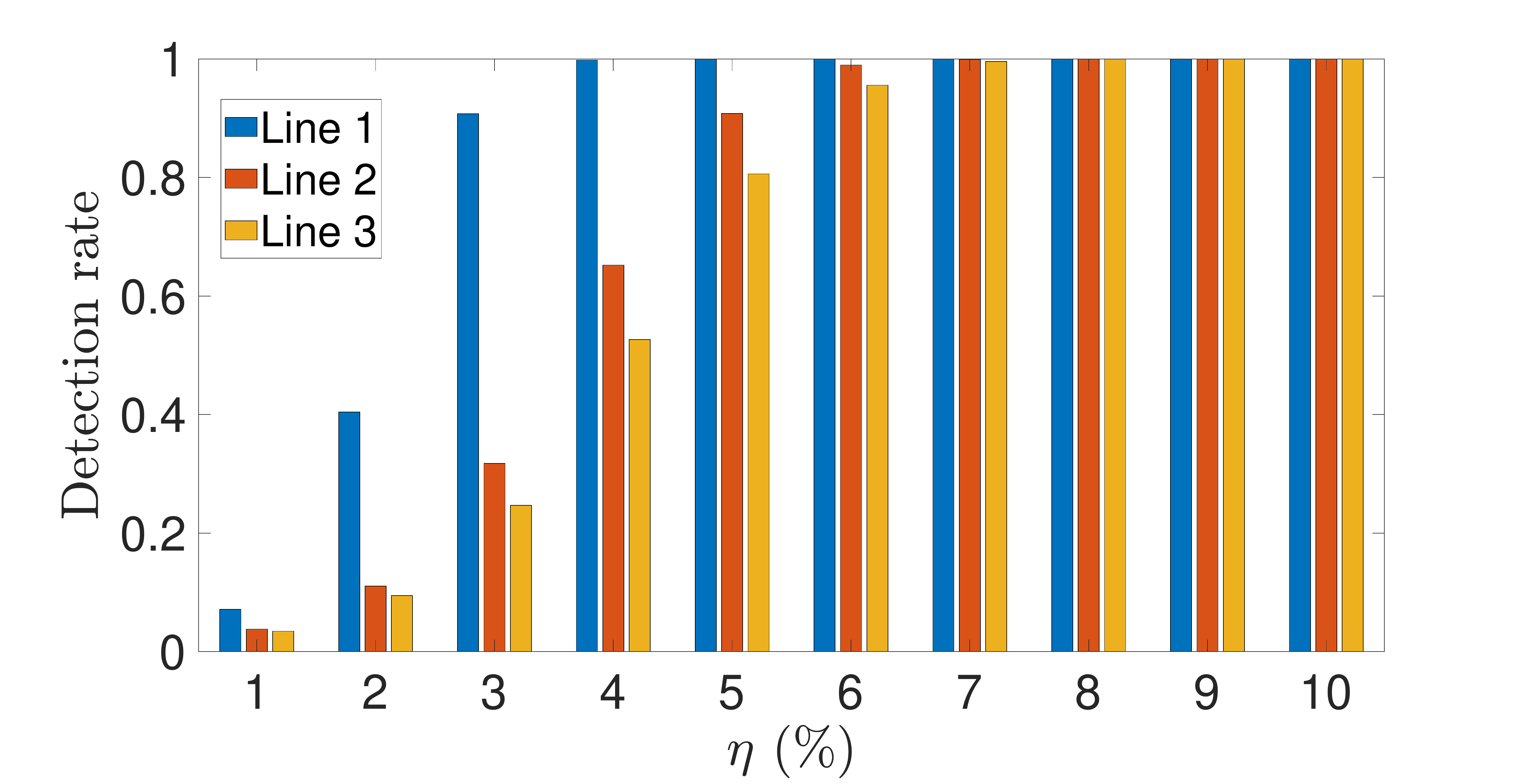}
\caption{Attack detection probability as a function of the percentage change ($\eta$) in the link reactance.}
\label{fig:Result1}
\end{figure}

First, we examine the D-FACTS deployment set problem. We perform simulations using the IEEE-14 bus system. 
As proposed in Section~\ref{sec:MTD_Soln}-A, we solve the minimum weight feedback edge set problem for the graph corresponding to the IEEE-14 bus system. Following this approach, the D-FACTS deployment set is given by $\mathcal{L}_D =  \{ 1,3,5,8,9,18,19 \}.$ We then perturb the reactances of all the links in the set  $\mathcal{L}_D.$ We simulate physical attacks against the three most important links in the system, i.e., Links 1, 2 and 3 (which have  the maximum power flow among all the links in the bus system) by disconnecting the links (one at a time), and injecting a corresponding CCPA of the form $\av = \Delta \Hm \thetav_p,$ where both $\Delta \Hm$ and $\thetav_p$ are computed using outdated knowledge of the system. We plot the BDD's attack detection probability for each case  in Fig.~\ref{fig:Result1} as a function of the percentage change in line reactances. It can be observed that the CCPAs can be detected with a high probability following the MTD approach. Moreover, about $5-6 \%$ perturbation in the line reactances is sufficient to achieve a high detection rate.
We also enlist the size of the D-FACTS deployment set for different IEEE bus systems in Table~\ref{tbl:DFACT_Size}. It can be observed that the proposed algorithm enables the defender to protect all the links in the system with only a few D-FACTS devices. Moreover, this is also the minimum-sized D-FACTS deployment set that can detect any CCPA against the grid. From the table, we can also conclude that $|\mathcal{L}_D|$ depends on not just the size of the bus system, but also its actual topology (e.g., $|\mathcal{L}_D| = 15$ for the 24 bus system, where as $|\mathcal{L}_D| = 9$ for the 39-bus system).

\begin{table}[!t]
 \begin{center}
 \begin{tabular}{||c | c | c | c ||} 
 \hline
 Bus system & $|\mathcal{L}|$ & $|\mathcal{L}_D|$  \\ [0.5ex] 
 \hline\hline
 IEEE 9-bus system & 9 & 1  \\ 
 \hline
 IEEE 14-bus system & 20 & 7  \\
 \hline
 IEEE 24-bus system & 38 & 15  \\
 \hline
 IEEE 39-bus system & 36 & 8  \\
  [1ex] 
 \hline
\end{tabular}
\end{center}
\caption{Size of the D-FACTS deployment set $|\mathcal{L}_D|$ for different IEEE bus systems.}
\label{tbl:DFACT_Size}
\vspace{-0.4 cm}
\end{table}

Finally, we show the effectiveness of the game-theoretic approach in reducing the operator's defense cost. 
The simulations are done on a IEEE-14 bus system. The generation cost is assumed to be linear, i.e., $C_i (G_{i,t}) = c_i G_{i,t}.$ The generators' capacities at buses $1,2,3,6,8$ are $G_{\max} = 300,50,30,50,20$~MWs and $c_i = 20,30,40,50,35$~\$/MWh respectively. $\fv_{\max}$ is chosen to be $160$ MWs for link $1,$ and $60$ MWs for all other links.
We consider two load conditions: (i) a heavily loaded system, scenario~1 with the load values at Bus 1 to 14 given by $0, 21.7, 94.2, 47.8, 7.6, 11.2, 0, 0, 29.5, 9, 3.5, 6.1, 13.5, 14.9$ MWs respectively, and (ii) a lightly loaded system, scenario~2 with the load values at Bus 1 to 14 given by $0,100,94.2,47.8,30,11.2,0,0,0;0,0,0,0,0$~MWs respectively. 
We consider five MTD perturbation strategies for the defender, i.e., $d_1 = \{1\}, d_2 = \{1,3\}, d_3 = \{1,3,5\}, d_4 = \{1,3,5,8\}, d_5 = \{1,3,5,8,9,18,19\}.$ We note that $d_5 = \mathcal{L}_D,$ which protects all the links of the system from CCPA. The attacker in turn launches a CCPA by disconnecting one of the links at a time. Under this set-up, we compute the NE solution in each of two scenarios  according to Definition~3 and the results are listed in Table~\ref{tbl:NE}.
It can be observed that the D-FACTS perturbation sets in the two scenarios are different. While, in the heavily loaded scenario (scenario~1), all the links in $\mathcal{L}_D$ need to be perturbed, in the lightly loaded scenario (scenario~2), it is sufficient to perturb only a subset of links. The rationale is that in the lightly loaded scenario, only a subset of links need to be protected from physical attacks, since the attacker is unlikely to target the unimportant links (i.e., the links that have very little power flow). We also list the defense cost, which is the percentage increase in the OPF cost. The NE solution of scenario~2 incurs much lower defense cost, since only a subset of links are perturbed. The above experiments show that the MTD perturbation set depends on the operational state of the system. By following the proposed game-theoretic approach, the operator can reduce its defense cost.

\begin{table}[!t]
 \begin{center}
 \begin{tabular}{||c | p{2cm} | c | c ||} 
 \hline
 Load scenario & NE D-FACTS perturbation set & Defense cost  \\ [0.5ex] 
 \hline\hline
 Scenario~1 & \{1,3,5,8,9,18,19\} & 11.62 \%  \\ 
 \hline
 Scenario~2 & \{1,3,5,8\} & 2.86 \%  \\
  [1ex] 
 \hline
\end{tabular}
\end{center}
\caption{D-FACTS perturbation set and the defense cost (the percentage increase in OPF cost) computed using the game-theoretic approach for different load scenarios.}
\label{tbl:NE}
\vspace{-0.4 cm}
\end{table}

\section{Conclusions and Future Work}
\label{sec:Conc}
In this work, we have proposed a novel strategy to detect CCPAs based on MTD and presented MTD design criteria in this context. We have identified the subset of links for D-FACTS device deployment that enables the defender to detect physical attacks against any link in the system. Further, to reduce the operator's defense cost, we have identified the set of links whose reactances must be perturbed at the operational time based on a game-theoretic approach.

There are still many open problems. First, D-FACTS devices are traditionally deployed in the grid with an objective of minimizing the transmission losses \cite{RogersDFACTS2008}. On the other hand, in this work, we discuss the D-FACTS device deployment problem from a security point of view only. These considerations suggest that the D-FACTS deployment problem will generally involve a trade-off between minimizing the transmission power losses and the security application.  
Another important problem arises in the game-theoretic formulation. 
Definition 3 provides a simple way to compute the mixed NEs by solving a system of linear
equations and checking some conditions. Still, in order to use it, one would have to know in advance the faces of
the simplex $\Delta_D \times \Delta_A$ on which the NE $(p_D^*, p_A^*)$ lies, i.e., one would
have to know the sets $\mathcal{I}^*_D$ and $\mathcal{I}^*_A$ for all NEs in advance. An exhaustive search has exponential complexity: the $(N_D+N_A)$-dimesional simplex has $2^{N_D+N_A}$ faces and all possibilities will have to be considered. Thus, a low-complexity algorithm must be found to compute the NE. We plan to investigate these issues in our future work. 

\balance
\bibliographystyle{IEEEtran}
\bibliography{IEEEabrv,bibliography}

\end{document}